# Logical Inferentialism & Attacks on Classical Logic

**Abstract**

This paper undertakes a foundational inquiry into logical inferentialism with particular emphasis on the normative standards it establishes and the implications these pose for classical logic. The central question addressed herein is: 'What is Logical Inferentialism & How do its Standards challenge Classical Logic?' In response, the study begins with a survey of the three principal proof systems that is, David Hilbert's axiomatic systems and Gerhard Gentzen's natural deduction and his sequent calculus, thus situating logical inferentialism within a broader proof-theoretic landscape. The investigation then turns to the core tenets of logical inferentialism by focusing on the role of introduction and elimination rules in determining the meaning of logical constants. Through this framework, natural deduction is evaluated as a system that satisfies key inferentialist virtues including harmony, conservativeness and the subformula property. Ultimately, the paper presents challenges to classical logic from intuitionist and revisionist perspectives by arguing that certain classical principles fail to uphold inferentialist standards, consequently undermining their legitimacy within a meaning-theoretic framework.

## 1 Introduction

To systematically investigate the relationship between logical inferentialism and classical logic plus examining how inferentialist standards challenge the consistency and legitimacy of classical reasoning, in section 2 we begin with a detailed survey of the three principal proof systems in logic namely, Hilbert's axiomatic systems and Gentzen's natural deduction and his sequent calculus. This foundational overview situates logical inferentialism within a broader proof-theoretic landscape and highlights the formal mechanisms by which each system models logical consequence. In section 3, we turn to the inferentialist theory of meaning which maintains that the content of logical constants is determined by their role in inference. This section traces the roots of logical inferentialism to Gentzen's remarks on the definitional function of introduction rules by arguing that inferential roles provide the semantic foundation for logical constants within natural deduction systems. Following this, section 4 evaluates Gentzen's natural deduction as a harmonious and well-structured proof system. Particular attention is given to the inferentialist virtues of harmony, normalisation, conservativeness, separability and the subformula property with an emphasis on how these virtues contribute to the internal lucidity of natural deduction and its status as a meaning-theoretic framework. The discussion also addresses Arthur Prior's tonk connective to illustrate how violations of harmony undermine logical legitimacy. Section 5 then presents a critique of classical natural deduction from the perspective of intuitionist and revisionist logicians such as Michael Dummett, Dag Prawitz and Neil Tennant. Their objections centre on classical principles most notably double negation elimination (DNE) and the law of excluded middle (LEM) which they argue violate inferentialist standards and compromise the transparency of semantic content. In section 6, we explore several responses from within the classical tradition, aimed at reconciling classical logic with inferentialist demands. These include Gentzen's sequent calculus, bilateral natural deduction systems such as Ian Rumfitt's and Greg Restall's interpretation of Michel Parigot's λμ-Calculus.[1] Each of these frameworks provides an alternative formulation of classical reasoning that preserves key inferentialist virtues while retaining classical expressiveness. While these responses demonstrate that classical logic is not immune to revision, they also suggest that inferentialism need not entail its outright rejection. The final section concludes by reflecting on the broader implications of this inquiry for both proof theory and the philosophy of logic by proposing that inferentialism, far from dismantling classical logic, may instead guide its refinement into more structurally disciplined and semantically transparent forms.

## 2 The Three Chief Proof Systems

Proof-theoretically speaking, a proof system consists of a formal language alongside a system of axioms and rules for performing logical inferences. To serve such a system, a logical language is typically characterised by a set of inductive clauses that define well-formed formulae.[2] The underlying schema assumes that the expressions of a formal language are particular sequences of symbols from a specified alphabet, constructed via inductive definitions. In this context, the expressions of a logical language are inductively defined formulae typically by means of two clauses. First, there are atomic or prime formulae i.e. those that contain no other formulae as constituents. Second, there are compound formulae which are constructed from more basic formulae through the use of logical constants. The meaning of compound formulae is determined by reference to the constituent formulae and the manner in which they are composed with the relevant logical symbols. Given any formula, one can determine how it was composed from other formulae and a logical constant. In many cases, parentheses are required to unambiguously indicate the structure of composition. By successively analysing the components in this way, one ultimately arrives at the atomic formulae. Consequently, all well-formed formulae are composed of atomic formulae, logical constants and parentheses. An alternative method for characterising formal languages involves the use of categorial grammars.[3] While such grammars are more commonly associated with natural languages, they are also employed in the formal analysis of programming languages. However, their application to logic remains relatively rare. In addition to a formal language, every proof system includes its own distinctive set of inference rules. One example of such a rule is that from $A \wedge B$ one may infer either A or B. Similarly, from $A \vee B$ it may be possible to infer a formula C, depending on the structure of the system. In a unilateral framework, inference rules apply to assertions alone. Conversely, in a bilateral construction, rules operate upon both assertions and denials, rather than solely on propositions.[4] For example, one can form an assertion from a proposition A by attaching a force marker such as 'It is the case that A' or 'It is not the case that A'.

[1] As discussed in section 6, Rumfitt (2000) presents a bilateral natural deduction system using signed formulae to distinguish assertion from denial, thus meeting inferentialist demands such as the principle of answerability, harmony, and normalisation. His co-ordination principles formalise reductio ad absurdum without relying on negation as assertion's dual. By contrast, Restall (2021) reformulates Parigot's λμ-Calculus through structural rules, and whether his system meets core inferentialist standards particularly answerability remains an open question.

[2] Proof theorists such as Sara Negri & Jan von Plato (2001) Page: 3 elucidate contemporary logical languages in the following words: *Logical languages of the present day have arisen as an abstraction from the informal language of mathematics. The first work in this direction was by Frege, who invented the language of predicate logic. It was meant to be, wrote Frege, "a formula language for pure thought, modeled upon that of arithmetic." Later Peano and Russell developed the symbolism further, with the aim of formalizing the language of mathematics. These pioneers of logic tried to give definitions of what logic is, how it differs from mathematics, and whether the latter is reducible to the former, or if it is perhaps the other way around. From a practical point of view there is a clear understanding of what logical languages are: The prime logical languages are those of propositional and predicate logic. Then there are lots of other logical languages more or less related to these. Logic itself is, from this point of view, what logicians study and develop. Any general definition of logic and logical languages should respect this situation.*

It is arguable that following Hilbert's discovery of proof theory in the early twentieth century numerous proof systems have been developed by various logicians. However, the most significant among them that is, those forming the foundation of all subsequent proof systems are Hilbert's axiomatic proof systems (commonly referred to as H-systems) and Gentzen's systems of natural deduction and sequent calculus (known respectively as N-systems and G-systems).[5] In Hilbert's axiomatic systems, in classical propositional calculus we begin with a set of axioms such as:[6]

[3] Joachim Lambek (1958) developed a sequent calculus for categorial grammar which interprets syntactic types as formulae and grammatical derivations as logical proofs. Lambek's system is noted for omitting structural rules such as weakening and contraction thus aligning closely with resource-sensitive logics like linear logic. This connection between logic and grammar highlights the structuralist thesis that the rules governing inference reflect not only logical truth but also syntactic form. In Lambek's calculus, for instance, word order and syntactic constraints are modelled using the non-commutative and non-associative structure of sequents.

[4] Negri & von Plato (2001) Page: 2 *There is a long-standing debate in philosophy on what exactly propositions are. When emphasis is on logic, and not on what logic in the end of a philosophical analysis is, one considers expressions in the formal sense and talks about formulas.*

[5] See Hilbert (1927), Hilbert & Wilhelm Ackermann (1928) and Gentzen (1935) cited in M. E. Szabo (1969).

[6] See Anne Sjerp Troelstra & Helmut Schwichtenberg (2000) Page: 51.

**Hc**

**Axioms**

| | | |
|---|---|---|
| **A → (B → A)** | **(A → (B → C)) → ((A → B) → (A → C))** | |
| **A → A ∨ B** | **B → A ∨ B** | **(A → C) → ((B → C) → (A ∨ B → C))** |
| **A ∧ B → A** | **A ∧ B → B** | **A → (B → (A ∧ B))** |
| **⊥ → A** | | |
| **A ∨ ¬A** | | |

which philosophically speaking, are tautologically true sentences. In Hilbert's system, when proving a formula e.g. A, any of the existing axioms or potentially a newly introduced one may be employed. In the case of propositional calculus, there is only one rule of inference namely, modus ponendo ponens or modus ponens (MP), also known as implication elimination:

$$\frac{A \rightarrow B \qquad A}{B}\,\text{MP}$$

In the case of predicate calculus, there is only one inference rule i.e. universal implication, or universal introduction:

$$\frac{A(x/y)}{\forall x A x}$$

With respect to assumptions, H-systems contain an important rule that is, in any proof regardless of its length, all assumptions must be open. In other words, there must be no closed assumptions whatsoever. However, with regard to N-systems, derivations begin without axioms and rely solely on assumptions and inference rules. The following are the inference rules of standard (textbook) classical natural deduction (NK) for propositional and predicate calculus:[7]

[7] There are numerous reformulations and presentations of natural deduction in the proof-theoretic literature. However, with some minor alterations, we shall adopt Gentzen's original formulation, as promoted by Prawitz since (1965). For the sake of notational simplicity, contextual formulae such as Γ and Δ have been omitted from these proofs. Nevertheless, it is understood that the initial formulae in the introduction rules are grounded in these contextual formulae. Furthermore, a significant matter regarding the presentation of predicate calculus' formalism should be stated namely, the distinction between bound or proper and free variables is crucial in this calculus. The formulae presented demonstrate how the variable y is bound or forms a proper formula, whereas t remains free. In the rule ∀I a variable must be arbitrarily chosen and not free in any undischarged assumption. The formula A(x/y) represents a generalisation where the y is a bound variable because its instantiation is arbitrary and does not depend on any particular term. When we conclude ∀xAx the variable x is now universally quantified i.e. it is formally bound within the scope of the quantifier. Conversely, in the rule ∀E the instantiation A(x/t) involves substituting the quantified variable x with a specific term t . However, t is not

**NK**

**Inference Rules**

**A B A ∧ B A ∧ B**
**________________∧I ______∧E ______∧E**
**A ∧ B A B**

**[A].[B]**
**Π**
**A ∧ B C**
**__________________________G∧E**
**C**

**[A] [B]**
**Π Π**
**A B A ∨ B C C**
**______∨I ______∨I _______________________________G∨E**
**A ∨ B A ∨ B C**

**[A]**
**Π**
**B A → B [A]**
**______________→I ________________→E**
**A → B B**

**[B]**
**Π**
**A → B A C**
**_______________________________G→E**
**C**

**⊥ ¬¬A**
**_____⊥E _____¬¬E**
**C A**

---

quantified within the proof making it a free term rather than a bound variable. Equally, in the rule ∃I replacing x with a particular term t allows us to assert that there exists an x for which Ax holds making t a free instance of the predicate. But in the rule ∃E a variable y is introduced as a bound variable to generalise the assumption A(x/y) ensuring that no dependency on any specific term t remains. Thus, y is bound when introduced arbitrarily in ∀I and ∃E, whereas t remains free when instantiated in ∀E and ∃I.

**[A]** **[A]** **[¬A]**
**Π** **Π** **Π**
**⊥** **A** **[¬A]** **C** **C**
**______¬I** **_________________¬E** **________________________Nem**
**¬A** **⊥** **C**

**[A(x/t)]**
**Π**
**A(x/y)** **∀xAx** **∀xAx** **C**
**______∀I** **______∀E** **______________________________G∀E**
**∀xAx** **A(x/t)** **C**

**[A(x/y)]**
**Π**
**A(x/t)** **∃xAx** **∃xAx** **C**
**______∃I** **______∃E** **______________________________G∃E**
**∃xAx** **A(x/y)** **C**

Finally, in G-systems axioms are used in place of assumptions. However, unlike Hilbert's axioms which are tautological formulae, the axioms of sequent calculus are themselves identity sequents e.g. A ⇒ A. In addition to the logical rules, which are presented as left and right rules, G-systems also contain structural rules such as weakening, contraction and cut. Among these three, the cut rule is considered the most fundamental principle of sequent calculus.[8] In sequent calculus, the application of the cut rule corresponds to the principle of compositionality in natural deduction:

**Γ ⇒ A** **A,Γ ⇒ C**
**__________________Subst**
**Γ ⇒ C**

specifically the process of combining two derivations in order to translate natural deduction proofs into sequent calculus derivations.[9] In such translations, the cut rule is necessary to represent elimination rules where the major premise is not an assumption but a derived formula. Crucially, it is required that the formula in the right premise of a cut be derived by a left-rule at some stage, as this mirrors the elimination rules of natural deduction in which an assumption is decomposed into its constituent components to be put to effective use. Furthermore, in the context of cut-free derivations, an inspection of the logical rules in sequent calculus reveals that every formula present in the premises of a logical rule also appears in its conclusion. This analytical feature is known as the subformula property. However, the presence of the cut rule undermines this property, meaning it no longer holds as a necessary condition. This is why proof theorists working in structural proof theory aim to develop systems of sequent calculus in which the cut rule is either eliminable or admissible.[10]

[8] The informal application of the cut rule can be observed in mathematical practice where proofs are divided into lemmas that can later be employed to derive further results.
[9] See Negri & von Plato (2001) Theorem 8.1.4 Page: 171.

In sequent calculus, we may transition from LK:[11]

**LK**

**Axioms**

$$A \Rightarrow A \qquad Ax$$

**Logical Rules**

$$\frac{A,\Gamma \Rightarrow \Delta}{A \wedge B,\Gamma \Rightarrow \Delta}\,L\wedge \qquad \frac{B,\Gamma \Rightarrow \Delta}{A \wedge B,\Gamma \Rightarrow \Delta}\,L\wedge \qquad \frac{\Gamma \Rightarrow \Delta,A \quad \Gamma \Rightarrow \Delta,B}{\Gamma \Rightarrow \Delta,A \wedge B}\,R\wedge$$

$$\frac{A,\Gamma \Rightarrow \Delta \quad B,\Gamma \Rightarrow \Delta}{A \vee B,\Gamma \Rightarrow \Delta}\,L\vee \qquad \frac{\Gamma \Rightarrow \Delta,A}{\Gamma \Rightarrow \Delta,A \vee B}\,R\vee \qquad \frac{\Gamma \Rightarrow \Delta,B}{\Gamma \Rightarrow \Delta,A \vee B}\,R\vee$$

$$\frac{\Gamma \Rightarrow \Delta,A \quad B.\Gamma \Rightarrow \Delta}{A \rightarrow B.\Gamma \Rightarrow \Delta}\,L\rightarrow \qquad \frac{A.\Gamma \Rightarrow \Delta,B}{\Gamma \Rightarrow \Delta,A \rightarrow B}\,R\rightarrow$$

$$\frac{\Gamma \Rightarrow \Delta,A}{\neg A,\Gamma \Rightarrow \Delta}\,L\neg \qquad \frac{A,\Gamma \Rightarrow \Delta}{\Gamma \Rightarrow \Delta,\neg A}\,R\neg$$

$$\frac{A(x/t),\Gamma \Rightarrow \Delta}{\forall xAx,\Gamma \Rightarrow \Delta}\,L\forall \qquad \frac{\Gamma \Rightarrow \Delta,A(x/y)}{\Gamma \Rightarrow \Delta,\forall xAx}\,R\forall$$

$$\frac{A(x/y),\Gamma \Rightarrow \Delta}{\exists xAx,\Gamma \Rightarrow \Delta}\,L\exists \qquad \frac{\Gamma \Rightarrow \Delta,A(x/t)}{\Gamma \Rightarrow \Delta,\exists xAx}\,R\exists$$

[10] See Negri & von Plato (2011).

[11] This formulation of LK with some minor modifications is adapted from Troelstra & Schwichtenberg (2000) Page: 61. For Gentzen's original formulation, see Gentzen (1935) cited in Szabo (1969) Page: 83.

**Structural Rules**

$$\frac{\Gamma \Rightarrow \Delta}{A,\Gamma \Rightarrow \Delta}\,LW \qquad \frac{\Gamma \Rightarrow \Delta}{\Gamma \Rightarrow \Delta,A}\,RW$$

$$\frac{A,A,\Gamma \Rightarrow \Delta}{A,\Gamma \Rightarrow \Delta}\,LC \qquad \frac{\Gamma \Rightarrow \Delta,A,A}{\Gamma \Rightarrow \Delta,A}\,RC$$

$$\frac{\Gamma \Rightarrow \Delta,A \qquad A,\Gamma \Rightarrow \Delta}{\Gamma \Rightarrow \Delta}\,Cut$$

namely, the classical multisuccedent sequent calculus (also referred to as G1c) to LJ i.e. the intuitionistic singlesuccedent sequent calculus (also known as G1i) simply by restricting LJ to at most one formula on the right-hand side of the ⇒.[12] However, at a more fundamental level, LK and LJ differ due to the inclusion or exclusion of classical principles such as LEM. More precisely, Gentzen's formulation of classical logic introduced the notion of a sequent in the form of a multisuccedent structure that is, expressions of the form $\Gamma \Rightarrow \Delta$ where both $\Gamma$ and $\Delta$ are multisets of formulae. Intuitionistic logic, in sequent calculus form, is obtained as a subsystem of classical logic by imposing restrictions on the contexts allowed in implication rules and, in predicate calculus, on the right-rule for the universal quantifier. In this manner, a consistent formalism for both intuitionistic and classical logic is achieved. Indeed, Gentzen's (1935) original LK was designed in such a way that the intuitionistic system LJ would emerge directly from it when the succedent is restricted to contain at most one formula. As a result, this proof system allows for sequents with empty antecedents that is, corresponding to derivations without assumptions in natural deduction as well as sequents with empty succedents corresponding to ⊥ (falsity) in natural deduction and even sequents where both antecedent and succedent are empty. In Gentzen's writings, the "denotational interpretation" of multisuccedent sequents refers to the reading of a sequent $\Gamma \Rightarrow \Delta$ as stating that the conjunction of the formulae in $\Gamma$ implies the disjunction of the formulae in $\Delta$. In his (1938) work, Gentzen describes the multisuccedent calculus as naturally reflecting the case distinctions commonly found in mathematical proofs. This is because the antecedent $\Gamma$ represents the open assumptions while the succedent $\Delta$ demonstrates the open cases of a derivation. For example, the logical rule L∧ conjoins two assumptions A and B into $A \wedge B$, while the rule R∨ combines two open cases A and B into $A \vee B$. If there is only one case, then the derivation has a singlesuccedent from the assumptions e.g. the sequent $\Gamma \Rightarrow C$. A sequent with no formula in the succedent such as $\Gamma \Rightarrow$ denotes an empty case i.e. a contradiction or impossibility. On the other hand, the singlesuccedent sequent $\Gamma \Rightarrow C$, as interpreted in natural deduction, expresses that the formula C is derivable from the assumptions in Γ. A further distinguishing feature of sequent calculus, in contrast to natural deduction, is its capacity to serve as a complete formal framework for classical logic including axiomatic, inferential and structural capabilities.[13] This strength is exemplified in Hacking (1979) who demonstrates that the system LK satisfies both the conservativeness principle and the subformula property namely, conditions which classical natural deduction (NK) fails to meet. Moreover, LK possesses an additional unique feature i.e. it satisfies the categoricity property which is not fulfilled by standard natural deduction systems. This was shown by Rudolf Carnap (1959) in Volume II of his *Introduction to Semantics and Formalization of Logic*.[14] Carnap proves that in a multisuccedent system such as LK, unlike in standard natural deduction, logical constants such as $\vee$ and $\neg$ are compatible not only with standard truth-functional semantics but also with certain non-standard (or deviant) interpretations.[15]

[12] Gentzen (1935) cited in Szabo (1969) Page: 86 states: *These basic sequents and our inference figures may easily be shown to be equivalent. The same possibility exists for the calculus LJ, with the exception of the inference figures ∨-IA and ¬-IS, since LJ-D-sequents may not in fact contain two S-formulae in the succedent (cf. V. 9 5).*

In summary, among the three proof systems discussed, Hilbert's axiomatic system possesses the admirable feature of having only a single rule of inference for both propositional and predicate calculus. However, in practice, Hilbert's method is rarely used due to the difficulty of identifying suitable axioms. Conversely, natural deduction closely resembles everyday reasoning and is widely used in mathematical proofs. Accordingly, no other proof system has attracted as much philosophical and theoretical attention as natural deduction. Nevertheless, sequent calculus systems enable the most profound investigations into the structure of proofs because sequent calculus functions as a formal theory of the derivability relation itself. As such, it is commonly employed in proof-theoretic analysis and in the study of consequence relations within a metalogical framework. A point of similarity between natural deduction and sequent calculus systems lies in the fact that their respective rule systems can be made equivalent that is, the same set of formulae can be derived in each. But the addition of the same rule to both systems may destroy this equivalence. In contrast, this lack of modularity does not arise in Hilbert-style axiomatic systems.

## 3 Logical Inferentialism

In the philosophy of language, more specifically, within the theory of meaning two central perspectives address linguistic meaning and mental content. These are referential (also known as denotational) approaches and use-theoretic approaches. According to the referential account, the semantic properties of linguistic expressions are primarily understood in terms of the referential relations they bear to extra-linguistic entities such as sets, numbers, proper names and other kinds of abstract or concrete objects. In this framework, linguistic content is determined by its correspondence with semantic values appropriate to its semantic category.

[13] See Ian Hacking (1979), Ray Cook (2005) cited in Stewart Shapiro (2005) and Negri & von Plato (2001).
[14] Detailed discussions about the categoricity problem can be found in Timothy Smiley & D. J. Shoesmith (1978), Hacking (1979), Nuel Belnap & Gerald Massey (1990), Smiley (1996), Rumfitt (1997 & 2000) and Lloyd Humberstone (2000).
[15] The categoricity problem and its link to natural deduction was originally discussed in Carnap (1943).

For example, proper names are taken to refer to specific objects, while basic predicates correspond to the attributes they ascribe. Thus, under a referential theory linguistic practice is explained in terms of such referential semantic relations. Conversely, use-theoretic accounts of meaning adopt an opposing standpoint. They propose that the regularities or rules governing linguistic use play a foundational role in explaining meaning and conceptual content. Semantic notions such as reference, truth, and satisfaction are thus understood as emerging from patterns of use rather than serving as primitive explanatory notions. It should be noted that use-theoretic theories of meaning exist in several distinct forms, differing primarily in which aspects of an expression's application are considered most fundamental. One well-known version of a use-based theory is inferentialism which holds that the content of mental states is determined by their inferential role namely, by the ways in which thoughts are used in reasoning and inference.

However a well-known and established form of inferentialism is logical inferentialism.[16] This form of inferentialism is rooted in Gentzen's (1935) doctoral thesis: *Investigations Into Logical Deduction* particularly his remarks that:

> The introductions represent, as it were, the 'definitions' of the symbols concerned, and the eliminations are no more, in the final analysis, than the consequences of these definitions. This fact may be expressed as follows: In eliminating a symbol, we may use the formula with whose terminal symbol we are dealing only 'in the sense afforded it by the introduction of that symbol.[17]

This well-known passage from Gentzen forms the foundation of logical inferentialism, as well as any theory of meaning grounded in natural deduction. This is because, according to Gentzen, in a natural deduction system introduction rules specify the necessary and sufficient conditions for establishing complex sentences, while elimination rules determine what may be justifiably inferred from them. For example, the meaning of conjunction is defined through its introduction rule:

$$\frac{\mathbf{A} \qquad \mathbf{B}}{\mathbf{A} \wedge \mathbf{B}} \wedge\mathbf{I}$$

By contrast, the consequence of using conjunction is understood through its elimination rules:[18]

[16] For a historical development of logical inferentialism, see William Kneale (1956), Hacking (1979), Michael Keremer (1988), Kosta Došen (1994), David Bostock (1997), Rumfitt (2000), Stephen Read (2000), Restall (2004, 2013, 2015 & 2021), Cook (2005), Florian Steinberger (2009) and David Ripley (2011, 2013 & 2017). It should be noted that although these authors did not contribute equally to the development of logical inferentialism, each played a significant role.

[17] Gentzen (1935) cited in Szabo (1969) Page: 80.

[18] Since ∧I and ∧E rules are truth-preserving, formulae of the form A ∧ B possess their standard truth-conditions.

| | |
|---|---|
| **A ∧ B** | **A ∧ B** |
| **______∧E** | **______∧E** |
| **A** | **B** |

Logical inferentialism maintains that meaning is determined by the inferential role of logical constants.[19] Accordingly, the inferential role of a logical constant is articulated through its introduction and elimination rules within a particular proof system i.e. natural deduction. Julian Murzi and Steinberger (2015) argue that, among all forms of inferentialism, logical inferentialism occupies a distinctive position. This is due to two characteristic features namely, inferentialist interpretations of logical vocabulary appear to be both natural and are typically regarded as the canonical form of inferentialism.[20] Consequently, logical inferentialism can be understood as a form of compositional theory of meaning in which the meanings of logical constants are determined by their introduction and elimination rules. However, although the meanings of logical constants are fixed by their inferential roles, logical inferentialism is not a conventionalist doctrine in which each individual is free to construct a logic of their own. As Steinberger (2011) notes, a logical inferentialist is expected to endorse those proof systems that reflect our shared inferential practices. This is because:

> After all, it is the practice represented, not the formalism as such, that confers meanings. Therefore, the formalism is of meaning-theoretic significance and hence of interest to the inferentialist only if it succeeds in capturing (in a perhaps idealised form) the relevant meaning-constituting features of our practice. It is in this sense, then, that the inferentialist position imposes strict demands on the form deductive systems may take. For future reference, let us refer to these demands as the Principle of answerability only such deductive systems are permissible as can be seen to be suitably connected to our ordinary deductive inferential practices.[21]

Arguably, the principle of answerability finds its clearest foundation in Gentzen's thesis that the meaning of logical constants is determined by their inferential role specifically through the interplay of introduction and elimination rules. In his (1935) dissertation, Gentzen maintains that introduction rules serve as definitional by establishing the sense of a constant, while elimination rules merely draw out the consequences already implicit in that introduction. This view implies that elimination rules derive their legitimacy from, and must remain subordinate to, the meaning fixed by introduction. Consequently, formal systems are not deemed meaningful simply because of their internal consistency; rather, they are accountable to, and constrained by, the inferential practices they aim to codify. On this inferentialist reading, the validity of a deductive system hinges on whether it reflects normatively governed reasoning patterns already operative in mathematical or discursive practice. It is precisely this requirement that Steinberger (2011) formalises under the principle of answerability, stipulating that only those systems which idealise shared inferential norms may be admitted as meaning-theoretically legitimate. In this sense, logical inferentialism resists both conventionalism and abstraction for its own sake, demanding that formal rules remain anchored in our common inferential commitments. By insisting on such accountability, the principle preserves the semantic transparency and normative grounding that Gentzen regarded as essential to logic's role in human reasoning. In addition to the requirement that our logical laws correspond to the principle of answerability, Dummett (1991) argues that, in order to avoid both conventionalism and circular reasoning, such laws must also be self-justifying.[22] This is because, if a logical law is regarded as basic or meaning-constitutive, we risk falling into the trap of Prior's (1960) notorious logical connective which will be discussed below.

[19] It is important to note that logical inferentialism defers from Robert Brandom's (1994) inferentialism which is a pragmatic view about mental content and meaning.
[20] For a discussion about the role and importance of logical vocabulary in inferentialism, see Brandom (2007).
[21] Steinberger (2011) Page: 335

## 4 Natural Deduction as a Harmonious Proof System

To illustrate the significance of natural deduction, Prawitz (1971) compares the power of Gentzen's natural deduction to Alan Turing's theory of computability since the former constitutes the ultimate proof system, while the latter represents the most fundamental mechanism in the theory of computation.[23] Additionally, owing to its closeness to human reasoning that is its naturalness, Gentzen's natural deduction systems have become the principal proof-theoretic framework in the semantic investigations of logical inferentialists. By contrast, axiomatic systems such as those developed by Gottlob Frege and Hilbert are far removed from the manner in which the human mind engages in reasoning. Gentzen characterises the naturalness of natural deduction in the following words:

> By means of a number of examples we shall first of all show what form deductions tend to take in practice and shall examine, for this purpose, three 'true formulae' and try to see their truth in the most natural way possible.[24]

In addition to examining the concept of logical consequence and the notion of logical validity, a natural deduction system offers a logical analysis of the inferential meaning of each logical constant. This analysis is conducted through the introduction and elimination inference rules. A significant feature of natural deduction is provided, that is it is carefully constructed, is the property of harmony. Broadly speaking, a natural deduction system is said to be harmonious if its elimination rules allow us to recover precisely the information that was originally introduced by its introduction rules.[25]

---

[22] Dummett (1991) Page: 245

[23] Prawitz (1971) Page: 246

[24] Gentzen (1935) cited in Szabo (1969) Page: 74. However, in personal correspondence via email, von Plato kindly informed me of the following quotation by stating that: *This is from Szabo's translation and it is simply wrong. Gentzen NEVER speaks about truth, and that is no accident. The original speaks about "richtig" and "Richtigkeit," that is, correct and correctness. Correctness is a clearly epistemic notion as are all notions you ever encounter in Gentzen's work.*

[25] Arguably, the concept of harmony is a contentious one. That is, there is no uniform agreement among philosophers of logic regarding the correct definition of this notion. Nevertheless, in order to gain some understanding of the various characterisations of the concept of harmony, one may refer to Read (2000) where a detailed account is provided of its development within proof-theoretic discussions of logical consequence. His analysis centres on key contributions by Gentzen, Prawitz, Dummett, and Hacking by highlighting both the philosophical commitments and the formal motivations that underpin the notion. In this regard, Read argues that the concept of harmony originates in a remark by Gentzen who described introduction rules in natural deduction as providing the definitions of logical constants with elimination rules serving as their consequences. This view

In natural deduction the introduction rule of an inference determines the meaning of the corresponding logical constant. Consequently, the elimination rules for that constant cannot be chosen arbitrarily and must be justified by reference to the matching introduction rule. This is because elimination rules are derivable from, and thus dependent upon, their corresponding introduction rules. For this reason, Dummett (1991) argues that introduction and elimination rules must be in harmony with one another.[26] Dummett's conception of harmony can be traced back to Prawitz's doctoral thesis *Natural Deduction: A Proof-Theoretical* Study (1965) as well as to Prawitz (1974) in which he characterises harmony in terms of the capacity to eliminate maximal formulae or "local peaks" that is, formulae that appear both as the conclusion of an introduction rule and as the major premise of the corresponding elimination rule.[27] For example, consider:

$$\dfrac{\dfrac{\mathbf{A} \qquad \mathbf{B}}{\mathbf{A \wedge B}}\,\wedge\mathrm{I}}{\mathbf{A}}\,\wedge\mathrm{E}$$

leads to a vision of logic as "autonomous" i.e. meaning is determined entirely by inference rules particularly through introduction rules. From this perspective, a logical constant is meaningful if its elimination rules do not permit one to derive more than is justified by its introduction rules. Prawitz formalised this idea through the process of normalisation by showing that if a formula is introduced and then immediately eliminated, this detour can be removed via simplifying the proof. This is often interpreted as ensuring that introduction and elimination rules are in harmony and the constant they govern is well-behaved. Dummett extended the notion of harmony by proposing that a logical constant is only acceptable if its use does not expand the inferential power of the system in undesirable ways. He distinguished between "intrinsic" (or "local") harmony which concerns the direct relationship between introduction and elimination rules and "total" harmony which includes broader constraints such as conservativeness namely, the requirement that adding the constant and its rules does not allow the derivation of new theorems in the original vocabulary. However, Read criticises Dummett's formulation of "total" harmony. He argues that conservativeness is too strong a constraint, potentially excluding well-behaved logical constants such as the truth predicate which can form non-conservative yet harmonious extensions. Furthermore, Read points out that Dummett's reliance on a so-called "fundamental assumption" i.e. any derivation can be transformed to end with an introduction rule, is invalid. He shows that this assumption fails in various cases including quantifiers and modal operators by leading to problematic conclusions. Hacking, by focusing on classical logic particularly negation, introduces further constraints that is, a logical constant must yield a Tarskian consequence relation namely, reflexive, monotonic and satisfying cut. He claims that if an extension violates these properties, it undermines the logic's structure. Yet, Read suggests that these conditions again might exclude legitimate constants and miss the more essential concern i.e. the relationship between rules and meaning. Ultimately, Read proposes that harmony should be understood primarily in terms of Gentzen's original insight that is, elimination rules must not exceed the content provided by the introduction rules. This relationship ensures that constants "are autonomous" with their meanings wholly determined by their inferential roles. He argues that this proof-theoretic notion of harmony when correctly interpreted can accommodate even classical connectives including negation within a lucid and self-justifying framework.

[26] Dummett (1991) Page: 246

[27] Prawitz (1965) Page: 34

In the above formalism, the conclusion A could be directly inferred from the two premises A and B. Consequently, the two ∧I and ∧E steps:

**________________∧I**
**A ∧ B**
**________________∧E**

are not necessary and are two detour steps. In other words, the formula A ∧ B is a maximum formula and can be eliminated. The normalised form of this construction is:

**A B**
**________________**
**A**

Although Dummett and Prawitz agree about the importance of the notion of harmony, to some extent they differ in their technical language. An example of this distinction is when the reduction reduces the degree of complexity of a derivation that is, the number of occurrences of logical operators. Dummett (1991) identifies this as "intrinsic" harmony and states: 'We may continue to treat the eliminability of local peaks as a criterion for intrinsic harmony; this is a property solely of the rules governing the logical constant in question.'[28] However, Dummett claims that while intrinsic harmony avoids elimination rules being stronger than their matching introduction rules such as Prior's tonk, it does not eliminate the chance that they be too weak. Prior, in his (1960) challenging paper *The Runabout Inference Ticket,* introduces tonk as a new logical constant, and with it, he introduces the following introduction and elimination inference rules:[29]

| | |
|---|---|
| **A** | **A tonk B** |
| **__________________tonk-I** | **__________________tonk-E** |
| **A tonk B** | **A** |

The fundamental problem with tonk is that if the consequence relation is transitive, and if even one formula can be proved in system *L*, then any formula can be proved in system *L*. However, if logicality is characterised via the notion of harmony, therefore the logical inferentialist should not be concerned about tonk. This is because, the tonk rules are not harmonious. Consequently, they are unable to characterise a logical constant. A crucial result of harmony is the property of conservativeness which does not permit us to introduce new logical constants e.g. tonk into the system in order to add expressive power to the language. Thus, Belnap (1962) demonstrates that tonk rules are not conservative because they trivialise the consequence relation. This suggests they permit us to prove sentences that were not provable in the past. Furthermore, Belnap argues that in a formal system admissible rules should be the ones which yield conservative extensions of the system to which they may be added.[30] Equally, the condition of conservativeness corresponds to the constraint that an admissible formal system e.g. *L* be separable. This means every provable sentence or rule in the system possesses a proof that includes only structural rules or rules for the logical constants that appear in that sentence or rule. This property is called separability. To finish this section, we should be aware of a significant fact i.e. natural deduction has a tendency towards intuitionistic logic and is not assembled for classical logic.[31] This is why a number of logicians and philosophers have promoted intuitionistic logic over classical logic and have taken a revisionist view towards our logical practice. As a result, in the next section we shall examine such ideas and critically discuss their benefits in accord with logical inferentialism.

[28] Dummett (1991) Page: 250
[29] Prior (1960) Page: 38

## 5 Revisionists' & Intuitionists' Attacks on Classical Logic

Intuitionists like Prawitz, Dummett and Tennant argue that although proof-theoretic notions such as harmony, conservativeness, and separability can overcome the dilemmas which have been put forward by tonk, standard classical systems, e.g. classical natural deduction, are not protected from such attacks.[32] These authors take an anti-realist position and attack semantic realism and classical logic from an intuitionistic logic standpoint. They argue that since in natural deduction the meanings of logical constants are determined via their introduction and elimination rules, therefore any practical theory of meaning must submit to these rules. Additionally, they believe that in order for our languages to be learnable, the inferential rules elucidating the meanings of the logical constants must possess the ability to be formalised and finite. Consequently, inferential rules must satisfy a number of properties, e.g. logical constants must be semantically well-behaved in order to fulfil the properties of harmony, normalisation, conservativeness, separability, categoricity and the subformula property. Proof-theoretically speaking, classical natural deduction, i.e. NK, is achieved by adding one of the classical laws, e.g. LEM, reductio ad absurdum (RAA), classical dilemma (CD) or DNE to the intuitionistic natural deduction NJ.[33] With respect to certain logical constants, we are aware that NK does not behave well and this is why opponents of classical logic argue that standard classical natural deduction cannot match intuitionistic natural deduction. The best-known advocate of this view is Dummett (1991) who believes that in a use-meaning approach concerning semantics, we must free ourselves from classical logic in favour of intuitionistic reasoning. As a result, intuitionists and revisionists subscribe to a proof-theoretic form of intuitionistic logic and use NJ as their preferred deductive system.[34]

[30] Belnap (1962) Page: 132

[31] At the same time that he acknowledges its power, Gentzen is aware of the shortcomings of natural deduction in proving laws of classical logic. Therefore in Gentzen (1935) cited in Szabo (1969) Page: 80,Gentzen writes: *5.1. The calculus NJ lacks a certain formal elegance. This has to be put against the following advantages: 5.11. A close affinity to actual reasoning, which had been our fundamental aim in setting up the calculus.*

[32] Prawitz, Dummett and Tennant hold since classical rules e.g. DNE are not harmonious and therefore not logical, the logical rules we should accept are those of intuitionistic logic that is, classical logic excluding the LEM principle and other classical equivalent rules. For a detailed proof-theoretic discussion about the vulnerability of classical logic, see Prawitz (1977), Dummett (1991) & Tennant (1997).

[33] In this work, we have named the classic proof-theoretic method known as CD as Nem. This is because, CD is actually the natural deduction form of LEM for arbitrary formulae.

A well-known classical proof-theoretic rule which cannot pass the intuitionists expectations is DNE:

$$\dfrac{\neg\neg \mathbf{A}}{\mathbf{A}}\ \neg\neg\mathbf{E}$$

This is because we are unable to prove the formula A from ¬¬A since the rule ¬E:

$$\dfrac{\mathbf{A} \qquad\qquad [\neg \mathbf{A}]}{\bot}\ \neg\mathbf{E}$$

is not in harmony with the rule ¬I:[35]

$$\dfrac{\begin{array}{c}[\mathbf{A}]\\ \mathbf{\Pi}\\ \bot\end{array}}{\neg \mathbf{A}}\ \neg\mathbf{I}$$

Furthermore, similar to ¬I, ¬E is not harmonious as it does not satisfy the subformula property. Correspondingly, standard axiomatisations of classical logic do not satisfy the notion of separability. For example, classical theorems such as Peirce's law ((A → B) → A) → A are provable via ¬E, ⊥E, ¬¬E and → and not solely by →. Similarly, the rules for ∨ and → are not sufficient to prove a classical theorem such as A ∨ (A → B). Equally, ¬E is unable to derive the formula A from ¬¬A and nor do we possess a well-behaved proof of LEM i.e. A ∨ ¬A. Although all these classical theorems are provable in classical natural deduction, this provability comes at the cost of losing the properties of harmony, normalisation, conservativeness, separability, categoricity and the subformula property or in other words, of not possessing a well-behaved proof system. To support some of our claims and then close this section, we shall respectively prove ((A → B) → A) → A, A ∨ (A → B) and A ∨ ¬A by the application of classical natural deduction in the following manner:

[34] Tennant (1987 & 1997) argues in our revisionist enterprise, we must employ a weaker logic even than intuitionistic logic i.e. intuitionistic relevant logic.

[35] This lack of harmony constitutes a meaning-theoretic failure because it allows for the derivation of consequences such as in the case of DNE that are not justified by the connective's introduction rule, thus obscuring the inferential role that defines the logical constant and undermining the transparency that a meaning-constitutive rule system is meant to preserve; in effect, the elimination rule extends the use of the constant beyond what its introduction rule warrants, severing the connection between syntax and semantic content that logical inferentialism aims to uphold.

**[A]  [¬A]**

**―――――― ¬E**

**⊥**

**―――――― ⊥E**

**B**

**―――――― →I**

**[(A → B) → A]        A → B**

**―――――――――――――――――――― →E**

**[¬A]        A**

**―――――――――――― ¬E**

**⊥**

**―――――――― ⊥E**

**¬¬A**

**―――――――― ¬¬E**

**A**

**―――――――――――――――――― →I**

**((A → B) → A) → A**

Note that Peirce's law involves the application of ¬¬E. Making use of DNE demonstrates that classical theorems such as Peirce's law are not analytic. This means that they do not simply follow from the semantic rules inherent in the constant → or from the rules of minimal or intuitionistic logic. Rather, they require additional, but not well-behaved, rules like ¬E, ⊥E and ¬¬E in order to derive the conclusion ((A → B) → A) → A. The formal that is, the sequent calculus form of  Peirce's law can be exhibited as:

**A ⇒ A**

**―――――――――――――――――― RW**

**A ⇒ A,B**

**―――――――――――――――――― L→**

**A → B ⇒ A**

**―――――――――――――――――― Ref**

**A → B ⇒ A    A ⇒ A**

**―――――――――――――――――― LC**

**(A → B),A ⇒ A**

**―――――――――――――――――― L→**

**(A → B) → A ⇒ A**

**―――――――――――――――――― R→**

**⇒ ((A → B) → A) → A**

Similarly, to derive A ∨ (A → B) as:

$$\frac{\dfrac{\dfrac{\dfrac{[A] \qquad [\neg A]}{\bot}\neg E}{B}\bot E}{A \rightarrow B}\rightarrow I}{A \vee (A \rightarrow B)}\vee I$$

in addition to ∨ and →, we also require the presence of ¬E and ⊥E. The formal presentation of A ∨ (A → B) can be demonstrated as:

$$\frac{\dfrac{\dfrac{A \Rightarrow A}{A \Rightarrow A, B}RW}{\Rightarrow A, (A \rightarrow B)}R\rightarrow}{\Rightarrow A \vee (A \rightarrow B)}R\vee$$

Ultimately, LEM can be derived as:

$$\frac{\dfrac{\dfrac{[\neg(A \vee \neg A)]}{\neg(A \vee \neg A) \vee (A \vee \neg A)}\vee I \qquad [\neg(\neg(A \vee \neg A) \vee (A \vee \neg A))]}{\bot}\neg E}{A \vee \neg A}\neg I$$

Like the derivation of ((A → B) → A) → A and A ∨ (A → B), the proof of A ∨ ¬A requires more than the already available connectives namely, ∨. To complete this proof we require extra rules such as ¬E and ¬I. The formal production of A ∨ ¬A can be produced as:

$$\frac{\dfrac{A \Rightarrow A}{\Rightarrow A, \neg A}R\neg}{\Rightarrow A \vee \neg A}R\vee$$

## 6 A Number of Possible Responses from Classical Logic

Although revisionist and intuitionist critiques of classical logic have mounted compelling objections grounded in logical inferentialism, they do not entail that classical logic is without recourse. A number of well-behaved proof systems have emerged within the classical tradition that explicitly respond to these inferentialist demands particularly the principles of harmony, normalisation, conservativeness, separability, and the subformula property. In response to the perceived inadequacies of classical natural deduction, these systems provide alternative formulations of classical reasoning that better accord with inferentialist standards. One notable example is Gentzen's sequent calculus particularly his LK which satisfies many of the criteria esteemed by inferentialists. Although LK is not a natural deduction system, its cut-elimination theorem and structural rules allow for a disciplined and well-behaved classical logic. The system avoids many of the shortcomings of classical natural deduction such as violations of separability and conservativeness by treating derivability as a relation between sets of formulae, rather than as mere syntactic deduction. Thus, while the system's multisuccedent nature may render it less intuitive from an inferentialist perspective, it nonetheless constitutes a formalism in which classical logic achieves a high degree of internal unity and proof-theoretic rigour.

Beyond sequent calculus, an influential development comes from the realm of bilateralism e.g. Rumfitt's system of bilateral natural deduction which for future reference, we label it as RBN:[36]

**RBN**

**Inference Rules**

| | | | |
|---|---|---|---|
| **+A** | **+B** | **+(A ∧ B)** | **+(A ∧ B)** |
| **________________+∧I** | | **_________+∧E** | **__________+∧E** |
| **+(A ∧ B)** | | **+A** | **+B** |

| | | | | |
|---|---|---|---|---|
| | | | **[-A]** | **[-B]** |
| | | | **Π** | **Π** |
| **-A** | **-B** | **-(A ∧ B)** | **C** | **C** |
| **______-∧I** | **______-∧I** | **_________________________________-G∧E** | | |
| **-(A ∧ B)** | **-(A ∧ B)** | **C** | | |
| | | | **[+A]** | **[+B]** |
| | | | **Π** | **Π** |
| **+A** | **+B** | **+(A ∨ B)** | **C** | **C** |
| **______+∨I** | **______+∨I** | **_________________________________+G∨E** | | |
| **+(A ∨ B)** | **+(A ∨ B)** | **C** | | |

[36] To fully read about these formalism, see Rumfitt (2000) Pages: 800 & 804.

$$\dfrac{-A \qquad -B}{-(A \vee B)}\;-\vee I \qquad \dfrac{-(A \vee B)}{-A}\;-\vee E \qquad \dfrac{-(A \vee B)}{-B}\;-\vee E$$

$$\dfrac{\begin{array}{c}[+A]\\ \Pi \\ +B\end{array}}{+(A \rightarrow B)}\;+\!\rightarrow I \qquad \dfrac{+(A \rightarrow B) \qquad +A}{+B}\;+\!\rightarrow E$$

$$\dfrac{+A \qquad -B}{-(A \rightarrow B)}\;-\!\rightarrow I \qquad \dfrac{-(A \rightarrow B)}{+A}\;-\!\rightarrow E \qquad \dfrac{-(A \rightarrow B)}{-B}\;-\!\rightarrow E$$

$$\dfrac{-A}{+(\neg A)}\;+\neg I \qquad \dfrac{+(\neg A)}{-A}\;+\neg E \qquad \dfrac{+A}{-(\neg A)}\;-\neg I \qquad \dfrac{-(\neg A)}{+A}\;-\neg E$$

**Co-ordination Principles**

$$\dfrac{\begin{array}{c}[\alpha]\\ \Pi \\ \bot\end{array}}{\alpha^*} \qquad \dfrac{\alpha \qquad \alpha^*}{\bot}$$

presents a sophisticated alternative to standard classical natural deduction by introducing force markers that distinguish between assertion and denial. This innovation permits the formulation of classical logical constants in a manner that directly addresses inferentialist concerns particularly with respect to negation and disjunction. Rumfitt's system includes a bilateral structure wherein signed formulae reflect both affirmative and negative commitments within proofs. In doing so, it accommodates the principle of answerability by ensuring that deductive moves correspond to recognisable inferential practices. Particularly noteworthy is Rumfitt's deployment of co-ordination principles which establish structural parity between signs and ensure that classical reductio methods can be expressed within the framework of natural deduction. This bilateral system has been defended as harmonious and normalisable and while some have raised concerns regarding its interpretive complexity, it remains one of the most compelling proof-theoretic responses to the inferentialist critique of classical logic. The explicit tracking of assertion and denial provides to Rumfitt's system a degree of semantic clarity often lacking in unilateral systems by suggesting that classical logic can be inferentially disciplined without surrendering its expressive power. However, a notable challenge of Rumfitt's account lies in its reliance on non-standard initial sequents and its departure from traditional intuitionistic formulations. This raises concerns about whether the bilateral system preserves the canonical structure of reasoning or introduces complications that may appear artificial or overly technical.

A further promising avenue is offered by Restall's (2021) in his interpretation of Parigot's λμ-Calculus. This system constitutes a classical singleconclusion natural deduction framework that employs labelled or reserved formulae to model alternative commitments. Restall argues that this formulation preserves the essential structure of natural deduction while permitting classical principles such as DNE and LEM:[37]

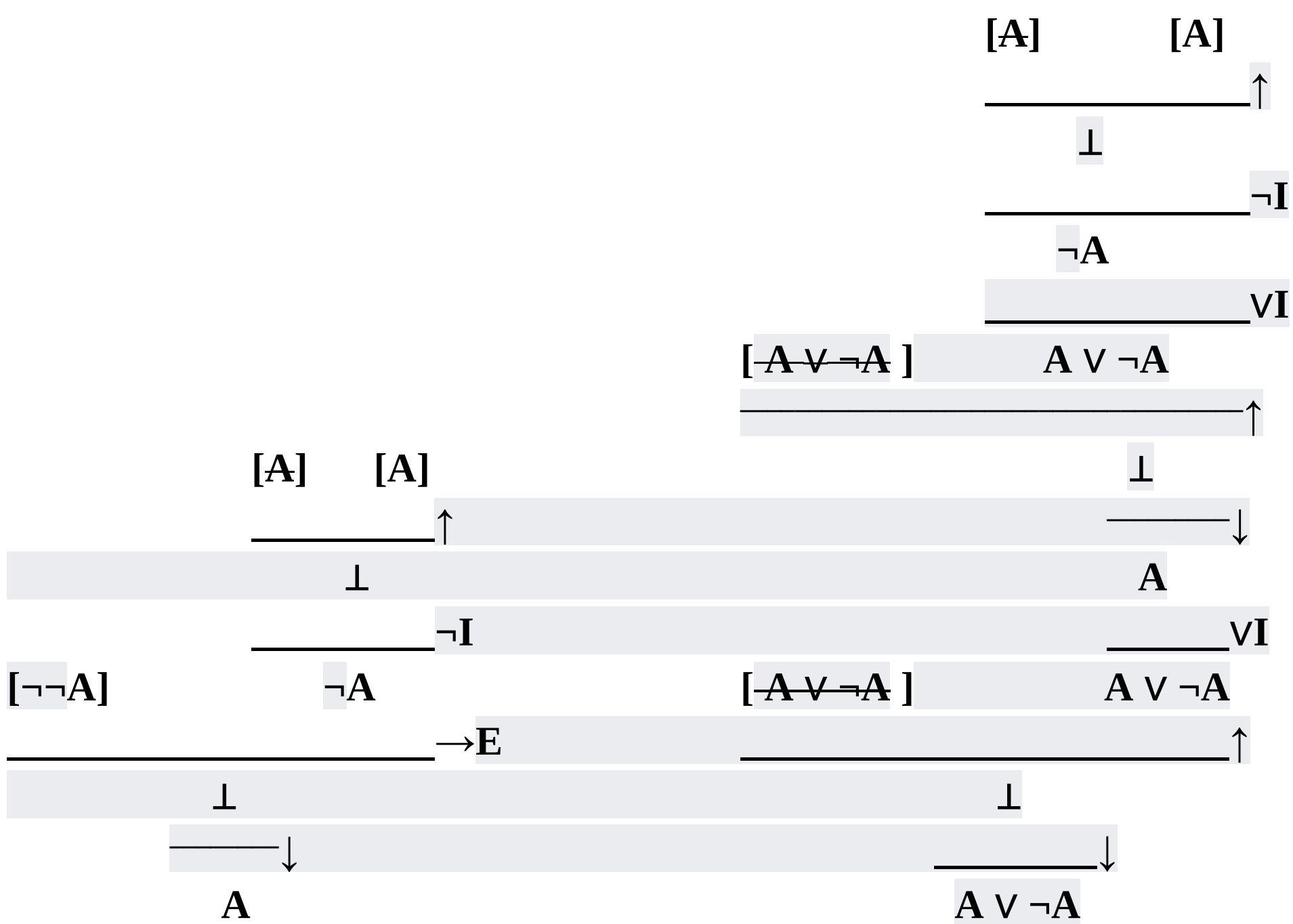

to be derived in a well-behaved manner. Unlike standard classical systems, Restall's understanding of λμ-Calculus integrates features of control operators from computational logic, thus reflecting a more nuanced understanding of how inferences unfold in structured discourse. Importantly, Restall's system aligns with the inferentialist aspiration to reflect the use of language in deductive practice by encoding assumptions and alternatives in the architecture of proof. In this framework, the speech act of asserting a conclusion is modulated by the reservation of alternatives, thus modelling the epistemic and dialogical aspects of reasoning that are often obscured in standard formalisms. This system, though technical, demonstrates that classical logic can be expressed in ways that do not violate inferentialist constraints, while retaining classical expressive completeness. Consequently, it offers a robust platform for further exploration into the compatibility of classical reasoning and inferentialist meaning theory. Nonetheless, in Restall's theory, a key challenge arises in the complexity of the structural machinery required. The λμ-Calculus introduces operational elements that, while technically powerful, may obscure the inferential meaning of the constants they are intended to define, thereby complicating their semantic transparency.

[37] Restall (2021) Page: 22

Collectively, these developments signal that classical logic is not bereft of avenues for rehabilitation in the face of inferentialist critique. Indeed, while standard textbook systems of classical natural deduction may fall short of the rigorous demands imposed by logical inferentialism, alternative formulations have demonstrated that the classical tradition possesses the internal resources to meet such challenges. Systems such as sequent calculus and Rumfitt's and Restall's systems exemplify classical frameworks that are harmonious, conservative and structurally consistent. These systems suggest that the logical inferentialist need not abandon classical logic wholesale, but can instead seek reformulations that retain classical strengths while embracing inferentialist virtues. However, to provide a detailed comparative analysis of these systems particularly in terms of their adherence to inferentialist standards and their potential to serve as meaning-theoretic foundations, would require an independent and dedicated study. Such an undertaking would allow for a systematic examination of their formal properties, their philosophical implications and their respective merits as models of logical consequence. As such, while this section acknowledges the existence and promise of these classical responses, a thorough assessment of their efficacy must be reserved for future work.

## 7 Concluding Remarks

The foregoing investigation has sought to clarify and evaluate the foundations, implications and challenges of logical inferentialism with a specific focus on its normative demands and their repercussions for classical logic. By charting the development of proof-theoretic systems from Hilbert's axiomatic frameworks to Gentzen's natural deduction and sequent calculi, the paper has situated logical inferentialism within a broader historical and technical context. Logical inferentialism, as articulated in this study, maintains that the meaning of logical constants is determined by their inferential roles, principally encoded through introduction and elimination rules. This perspective aligns most naturally with the structure of natural deduction systems where the form and force of inferential steps provide a transparent framework for meaning-constitutive acts. Central to this theoretical framework is the principle of harmony which demands that elimination rules mirror and remain justified by their corresponding introduction rules. The resulting conceptual structure imposes strict constraints on the admissibility of logical constants and inference rules, thus excluding deviant connectives such as Prior's tonk. Moreover, the inferentialist commitment to properties such as normalisation, conservativeness, separability and the subformula property not only reaffirms the foundational integrity of proof-theoretic systems but also serves as a standard by which classical logic is critically examined and frequently found wanting.

The challenge posed by intuitionist and revisionist logicians to classical natural deduction lies at the heart of the inferentialist critique. Pioneers such as Dummett, Prawitz and Tennant have demonstrated that a number of classical principles particularly DNE and LEM fail to satisfy the inferentialist conditions of harmony and separability. These failures render the classical system structurally inadequate from the perspective of meaning-theoretic transparency. For example, the inclusion of DNE within classical natural deduction violates the subformula property and relies on inference steps that cannot be derived solely from the introduction rules of negation. Likewise, classical theorems such as Peirce's law or LEM cannot be proven without invoking supplementary rules that are neither harmonious nor conservative. This deficiency results in a breakdown of analytic transparency whereby the inferential content of logical constants becomes opaque, thus undermining the notion that logical rules express inferential practices rather than syntactic artefacts. These concerns lend substantial weight to the anti-realist programme which regards intuitionistic logic as a more faithful formalisation of our inferential commitments. By insisting on structural virtues such as proof normalisation and the elimination of maximal formulae, the intuitionist position highlights classical logic's apparent inability to maintain unity between semantics and proof structure. Consequently, within the inferentialist framework, the classical tradition appears theoretically vulnerable and in need of systematic revision or replacement.

However, as the final sections of the paper have shown, classical logic is not without recourse in the face of these critiques. A number of alternative classical systems have emerged that demonstrate a capacity to meet the inferentialist's rigorous standards. Chief among these is Gentzen's sequent calculus particularly his LK which enjoys properties such as cut-elimination, subformula preservation and structural discipline. Though not strictly a natural deduction system, LK's proof-theoretic robustness and capacity to model derivability relations render it an inferentially acceptable candidate within the classical tradition. Additionally, bilateral natural deduction systems such as that developed by Rumfitt provide a promising strategy for retaining classical expressiveness while observing inferentialist constraints. By explicitly tracking assertion and denial through signed formulae, these systems restore harmony and answerability to classical negation and disjunction, thus resolving many of the concerns raised by intuitionists. Similarly, Restall's reinterpretation of Parigot's λμ-Calculus presents a structurally refined version of classical natural deduction that integrates alternative commitments while preserving inferential integrity. Taken together, these systems illustrate that classical logic can evolve in response to inferentialist critique by adopting new structural forms that retain its core principles while adhering to meaning-theoretic standards. In so doing, they reinvigorate the classical tradition through suggesting that inferentialism need not entail a wholesale rejection of classical logic but may instead serve as a constructive force for its refinement. Future research might further develop these alternative systems, assessing their comparative merits and limitations and exploring the broader philosophical implications of reconciling classical logic with inferentialist demands.

Beyond the confines of mathematical and philosophical logic, the issues explored in this paper hold significant implications for adjacent disciplines. In computer science particularly in areas such as programming language design, type theory, and formal verification, the inferential structure of logical systems plays a crucial role in determining how computation and reasoning interact. Systems grounded in harmonious and meaning-constitutive rules align well with proof-theoretic interpretations of computation, as seen in the Curry-Howard correspondence, where types are interpreted as propositions and programmes as proofs. A refined inferential understanding of logical constants can thus inform the development of more robust, semantically transparent programming frameworks. Likewise, in the philosophy of mathematics, the inferentialist approach invites a reconsideration of the foundations of mathematical reasoning. If classical logic is to be retained, it must be justified not simply by its utility but through its alignment with shared inferential norms. This raises critical questions for foundational programmes that rely on classical assumptions, such as set theory or model theory by challenging them to either reconstruct their logic inferentially or defend its meaning-theoretic lucidity. Furthermore, debates over logical harmony and structural rules resonate with ongoing discussions in mathematical practice about constructivism, intuitionism, and pluralism. By situating logical inferentialism within this broader intellectual landscape, we begin to appreciate how the defence or revision of classical logic is not merely a technical or philosophical concern, but one that bears directly on the methods and conceptual frameworks of the sciences and mathematics at large.

## Acknowledgement

We are grateful to Florian Steinberger for reading an early draft of this paper and for offering constructive suggestions and remarks. Without his ongoing support, this paper would not have developed into its present form. We are also thankful to Jan von Plato for his careful attention to the manuscript, particularly for his correction of footnote (24) regarding Gentzen's translation. We further wish to acknowledge the anonymous referee who generously read the paper online and provided insightful comments on various sections. Finally, We are indebted to Birkbeck College, University of London, for providing the research facilities that made the writing of this work possible.